\documentclass[aps,prl,floatfix,superscriptaddress,reprint]{revtex4-1} 
\usepackage{graphicx}
\usepackage{amsmath,amsfonts,amssymb}
\usepackage{color}
\usepackage{pdfpages}
\usepackage{tabularx}
\usepackage{comment}

\makeatletter
\AtBeginDocument{\let\LS@rot\@undefined}
\makeatother

\usepackage{pdfpages}

\begin{document}

\title{Prediction of Double-Weyl Points in the Iron-Based Superconductor CaKFe$_4$As$_4$}
\date{\today}
\begin{abstract}
	 Employing a combination of symmetry analysis, low-energy modeling, and
	 {\it ab initio} simulations, we predict the presence of
	 magnetic-field-induced Weyl points close to the Fermi level in 
	  CaKFe$_4$As$_4$.
	 Depending on the relative strengths of the magnetic field and of the spin-orbit coupling, the Weyl fermions
	 can carry a topological charge of $\pm1$ or $\pm2$, 
	 making CaKFe$_4$As$_4$ a rare realization of a double-Weyl semimetal.
	 We further predict experimental manifestations of
          these Weyl points, both in bulk properties, such as the
         anomalous Hall effect, and in surface properties,
such as the emergence of prominent Fermi arcs. Because  CaKFe$_4$As$_4$
	 displays unconventional fully-gapped superconductivity below 30 K,
	 our findings open a novel route to investigate the 
	 interplay between superconductivity and Weyl fermions.
\end{abstract}

\author{Niclas Heinsdorf}
\email{heinsdorf@itp.uni-frankfurt.de}
\affiliation{Institut f\"ur Theoretische Physik, Goethe-Universit\"at Frankfurt,
Max-von-Laue-Strasse 1, 60438 Frankfurt am Main, Germany}
\author{Morten H. Christensen}
\altaffiliation{Present Address: Center for Quantum Devices, Niels Bohr Institute, University of Copenhagen, Copenhagen 2100, Denmark}
\affiliation{School of Physics and Astronomy, University of Minnesota, Minneapolis, 55455 MN}
\author{Mikel Iraola}
\affiliation{Institut f\"ur Theoretische Physik, Goethe-Universit\"at Frankfurt,
Max-von-Laue-Strasse 1, 60438 Frankfurt am Main, Germany}
\affiliation{Department of Condensed Matter Physics, University of the Basque Country UPV/EHU, Apartado 644,
48080 Bilbao, Spain}
\affiliation{Donostia International Physics Center, 20018 Donostia-San Sebastian, Spain}
\author{Shang-Shun Zhang}
\affiliation{Department of Physics and Astronomy, The University of Tennessee, Knoxville, Tennessee 37996, USA}
\author{Fan Yang}
\affiliation{Department of Chemical Engineering and Materials Science, University of Minnesota, Minneapolis, 55455 MN}
\author{Turan Birol}
\affiliation{Department of Chemical Engineering and Materials Science, University of Minnesota, Minneapolis, 55455 MN}
\author{Cristian D. Batista}
\email{cbatist2@utk.edu}
\affiliation{Department of Physics and Astronomy, The University of Tennessee, Knoxville, Tennessee 37996, USA}
\affiliation{Neutron Scattering Division and Shull-Wollan Center,
Oak Ridge National Laboratory, Oak Ridge, Tennessee 37831, USA}
\author{Roser Valent{\'\i}}
\email{valenti@itp.uni-frankfurt.de}
\affiliation{Institut f\"ur Theoretische Physik, Goethe-Universit\"at Frankfurt,
Max-von-Laue-Strasse 1, 60438 Frankfurt am Main, Germany}
\author{Rafael M. Fernandes}
\email{rfernand@umn.de}
\affiliation{School of Physics and Astronomy, University of Minnesota, Minneapolis, 55455 MN}

\maketitle

The realization of topological phenomena in iron-based superconductors has
opened a new route to elucidate the interplay between topology, electronic correlations,
and unconventional superconductivity. Indeed, a band inversion involving an
As/Se $p_z$-band and a pair of Fe $d_{xz}/d_{yz}$-bands along the $\Gamma$-$Z$
line of the Brillouin zone has been observed in several compounds, such as
FeTe$_{1-x}$Se$_x$, LiFe$_{1-x}$Co$_x$As, and CaKFe$_4$As$_4$~\cite{wang2015topological,zhang2019multiple,lohani2020band,borisenko2020strongly,liu2020a}. In the normal
state, such a band inversion gives rise to helical Dirac states at the surface
and semimetallic Dirac states in the bulk. In the superconducting state,
zero-energy states are observed inside some (but not all) vortices, suggesting
the presence of Majorana bound states~\cite{yin2015observation,wang2018evidence,liu2018robust,machida2019zero-energy,kong2020tunable}, whereas a flat density of states is seen
at domain walls, characteristic of linearly-dispersing one-dimensional Majorana
modes~\cite{wang2020evidence}.

In this paper, we show that another non-trivial topological phenomenon --
Weyl points -- can be realized in the iron-based superconductor
CaKFe$_4$As$_4$ in a magnetic field. The existence of these Weyl fermions, which are anchored in a fourfold rotational symmetry of the lattice, does not rely on a $p$-$d$ band inversion, but on
the fact that each Fe plane in this  bilayer compound lacks inversion symmetry. Unlike the other iron-based superconductors, the inversion centers of CaKFe$_4$As$_4$ are not on the Fe layer, but at the positions of the Ca and K atoms. Consequently, there is no glide-plane symmetry either.
While the inversion symmetries on the Fe layers are broken explicitly by the lattice, 
the effect is enhanced by magnetic fluctuations associated with the nearby
spin-vortex crystal state realized in weakly electron-doped CaKFe$_4$As$_4$~\cite{meier2018hedgehog}.

Our combined analysis involving symmetry considerations
and {\it ab initio} simulations demonstrates that, in the presence of a magnetic field, a pair of Weyl points emerge close to the Fermi energy along the high-symmetry $M$-$A$ $\left[ \textrm{i.e.} \, (\pi,\pi,0) \textrm{-} (\pi,\pi,\pi) \right]$ line of the Brillouin zone.
Whereas a number of 
inversion-symmetry-broken Weyl semimetals have been reported
\cite{huang2015weyl, weng2015weyl, xu2016observation, xu2015discovery,
xu2015discovery2, yang2015weyl, ruan2016symmetry, soluyanov2015type,
sun2015prediction, wang2016mote, chang2016prediction, chang2018magnetic,
xu2016discovery3}, time-reversal symmetry-broken Weyl semimetals seem more
scarce \cite{wan2011topological,borisenko2015time,
chinotti2016electrodynamic,chang2016room, wang2016time, hirschberger2016chiral,
shekhar2016observation, suzuki2016large, yang2017topological, zhang2017strong}. 
Importantly, in CaKFe$_4$As$_4$ we show that depending on the relative magnitudes of the splittings of energy levels at the $M$ point caused by the Zeeman field, by the bilayer coupling, and by the spin-orbit coupling, Weyl points can arise from crossings between bands of opposite or equal spin polarization. In the latter case, which we explicitly verify in our {\it ab initio} calculations, the Weyl fermions carry a higher order topological charge of $\pm2$, making CaKFe$_4$As$_4$ a {\it double}-Weyl semimetal. Experimentally, we  propose that the presence of Weyl fermions can be probed by transport measurements of the anomalous Hall effect and by spectroscopic detection of the characteristic Fermi arcs that we obtain in our analysis.

We propose that the separation between the Weyl points and the Fermi level is reduced not only by hole doping, but also by electronic correlations, which are known to generally shrink the bands in iron-based superconductors \cite{Ortenzi2009,yin2011kinetic,ferber2012,Chubukov2016,zantout2019,Bhattacharyya2020}. Moreover, since the orbitals from which the Weyl points originate are the same $d$-orbitals that become superconducting below $T_c \approx 30$ K, CaKFe$_4$As$_4$ provides a promising framework to realize an intrinsic unconventional superconducting double-Weyl semimetal.

\begin{figure}
    \centering
    \includegraphics[width=0.9\linewidth]{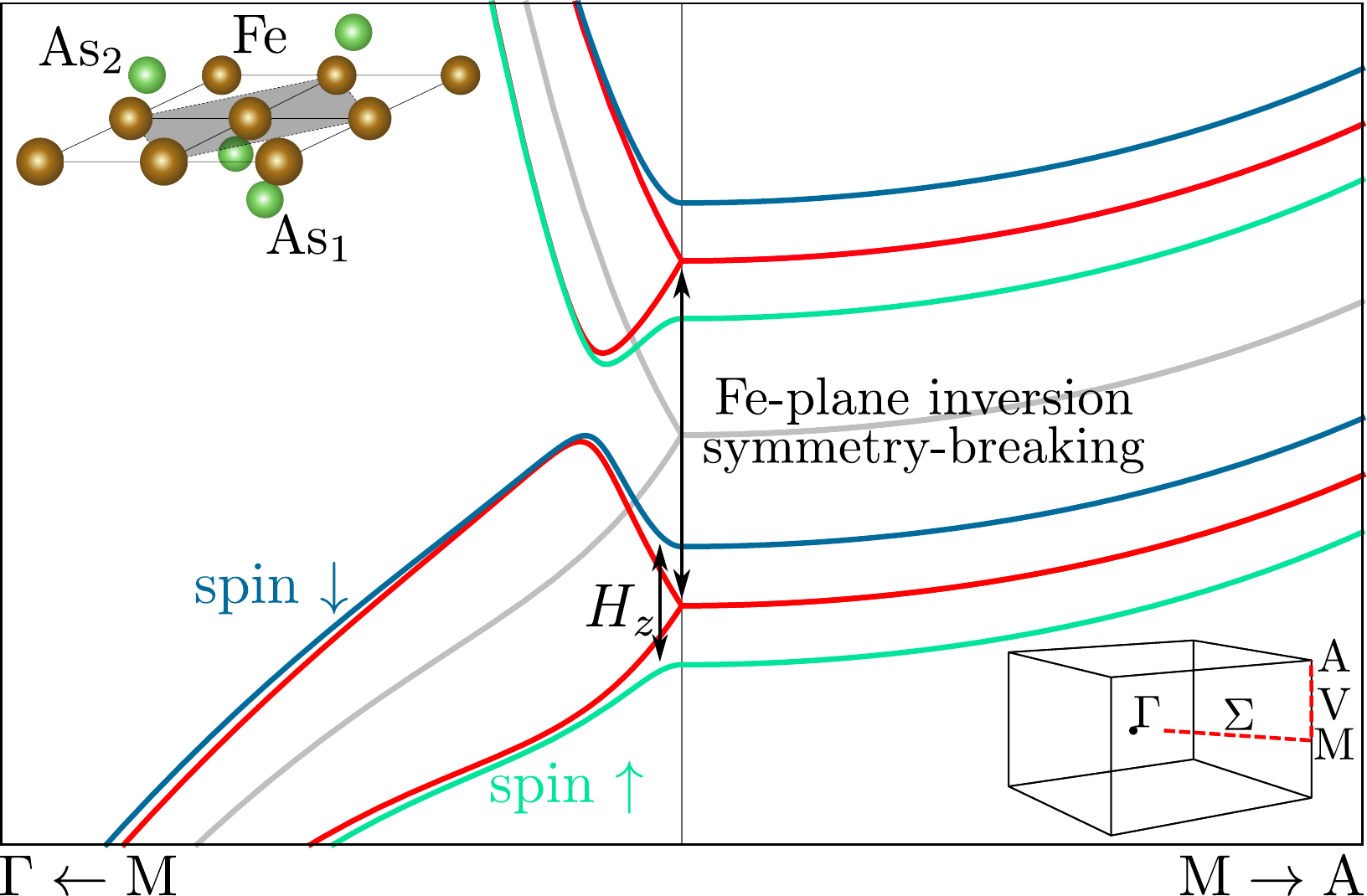}
    \caption{Schematic of the symmetries and band-dispersion degeneracies of a single FeAs layer with $P4/nmm$ space group. 
    Including spin-orbit coupling, the glide-plane symmetry enforces fourfold degenerate bands along the $M$-$A$ line and twofold degenerate bands along $\Gamma$-$M$ (gray curves). Breaking of the Fe-plane inversion symmetry by making the two As atoms above and below the plane inequivalent (upper inset) lowers the degeneracies of the bands to twofold along $M$-$A$ and one-fold along $\Gamma$-$M$ (red curves). Application of a perpendicular magnetic field makes all bands non-degenerate (green and blue curves). The lower inset shows the Brillouin zone, with $\Gamma =(0,0,0)$, $M =(\pi,\pi,0)$ and $A =(\pi,\pi,\pi)$. }
    \label{fig:Fe_plane}
\end{figure}

%
We first consider a single FeAs layer with $P4/nmm$ (\#129) space group, and
analyze which symmetries need to be broken in order to obtain non-degenerate bands, whose crossings can result in Weyl points. 
The system has an
inversion center on the Fe plane and, due to the puckering of the As atoms above and below the Fe plane, a glide-plane
symmetry consisting of a reflection with respect to the Fe plane followed by a
half unit-cell translation along the in-plane diagonal (upper inset of Fig.~\ref{fig:Fe_plane}).  The combination of this Fe-plane centered inversion symmetry  
with time-reversal symmetry leads to a Kramers degeneracy of the bands over the whole Brillouin zone, as illustrated in Fig.~\ref{fig:Fe_plane} by the twofold degenerate bands along the in-plane $\Gamma$-$M$ direction. 
The energy states at the $M$ point, however, are fourfold degenerate, even in the presence of spin-orbit
coupling (SOC), due to the glide-plane symmetry of the space group ~\cite{cvetkovic2013space, Hund1936} (gray curves in Fig. \ref{fig:Fe_plane}).

For a simple stacking of FeAs layers (i.e. a stacking that preserves the non-symmorphic $P4/nmm$ space group of a single layer, like in FeSe or LiFeAs), the band degeneracies at the $M$ point extend along the entire $M$-$A$ line.
When inversion symmetry centered on the Fe plane is broken so that the space group is reduced to $P4mm$, either explicitly by a substrate~\cite{Halloran2017} or spontaneously by interactions that favor a spin-vortex crystal magnetic ground state~\cite{meier2018hedgehog}, the glide-plane symmetry is also broken and some of these degeneracies are lifted. As shown by the red curves in Fig. \ref{fig:Fe_plane}, the fourfold degenerate bands along $M$-$A$ split into pairs of twofold degenerate bands in the presence of SOC. Along the in-plane $\Gamma$-$M$
line, all bands become non-degenerate, as previously discussed in Ref. \cite{christensen2019intertwined}. The remaining twofold degeneracy of the energy levels along $M$-$A$ can be lifted by the Zeeman coupling to an external out-of-plane magnetic field, as shown by the green and blue curves in Fig. \ref{fig:Fe_plane}. 
Depending on the band structure parameters and on the magnitude of the magnetic field, these non-degenerate bands can potentially cross, yielding a pair of Weyl points at
$(\pi,\pi,\pm k_z)$. 

\begin{figure}
    \centering
    \includegraphics[width=1.\linewidth]{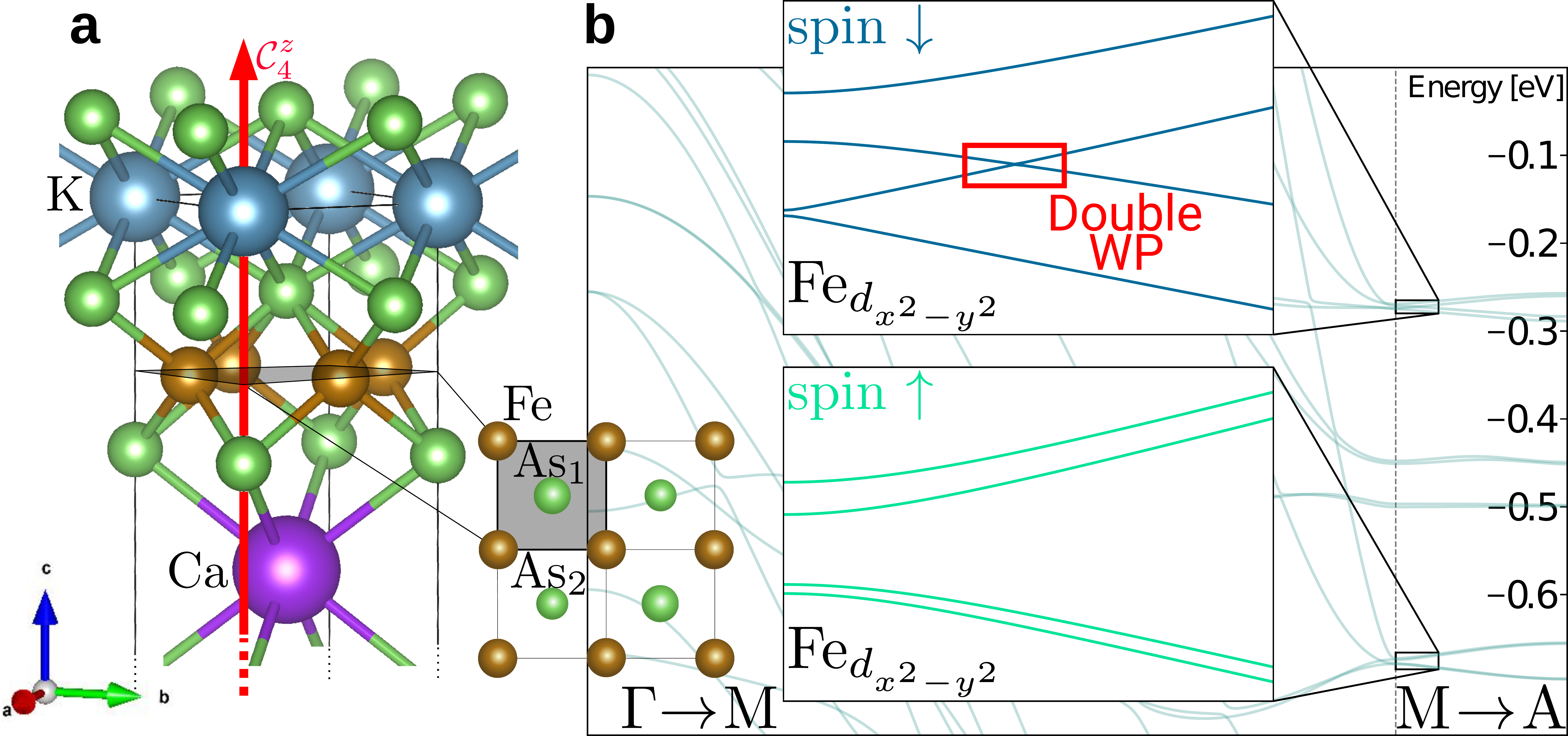}
    \caption{(a) Crystal structure of CaKFe$_4$As$_4$. (b) DFT band structure in the presence of finite Fe magnetization and SOC calculated with the FPLO basis \cite{koepernik1999full}. A double Weyl-point involving Fe $d_{x^2-y^2}$ orbitals with equal spin emerges.}
    \label{fig:cell_and_bands}
\end{figure}

Having established a general framework in which Weyl points can potentially arise in iron-based superconductors, we now turn our attention to a specific material. An ideal candidate to realize this effect is the bilayer CaKFe$_4$As$_4$, since the individual FeAs layers lack inversion symmetry [see Fig.~\ref{fig:cell_and_bands}(a)]. 
Indeed, this material crystallizes in the symmorphic tetragonal
space group $P4/mmm$ (\#123) with lattice constants 
$a$ = $b$ = 3.8659~\AA\ and $c$=12.884~\AA. The fourfold rotational symmetry about the $z$-axis, $C_4^z$, is centered at either the middle of the conventional unit cell or at one of its four corners. The distinguishing feature of this crystalline structure is that the alternating layers of Ca and K make the two As atoms inequivalent, thus breaking the glide plane symmetry present in all other iron-based superconductors. This results in a polar structure for the Fe layers, since the site-symmetry of the Fe atoms is $2mm$ ($C_{2v}$), with no mirror plane or center of inversion on the Fe layer. Moreover, because CaKFe$_4$As$_4$ is at the verge of spin-vortex crystal order~\cite{meier2018hedgehog}, magnetic fluctuations are expected to enhance the impact of the explicit glide-plane symmetry-breaking on the low-energy electronic states.

In contrast to the single-layer case discussed in the context of Fig. \ref{fig:Fe_plane}, the unit cell of CaKFe$_4$As$_4$ contains two Fe layers. This ensures the existence of an inversion center on the Ca and K sites, which in turn preserves the Kramers degeneracy of the bands for every momentum.  
To verify whether Weyl points can still emerge for this bilayer stacking configuration, we first perform a group-theoretical analysis appropriate for the space group $P4/mmm$, focusing on the $\Gamma$-$M$-$A$ path.

\begin{figure}
    \centering
    \includegraphics[width=1.\linewidth]{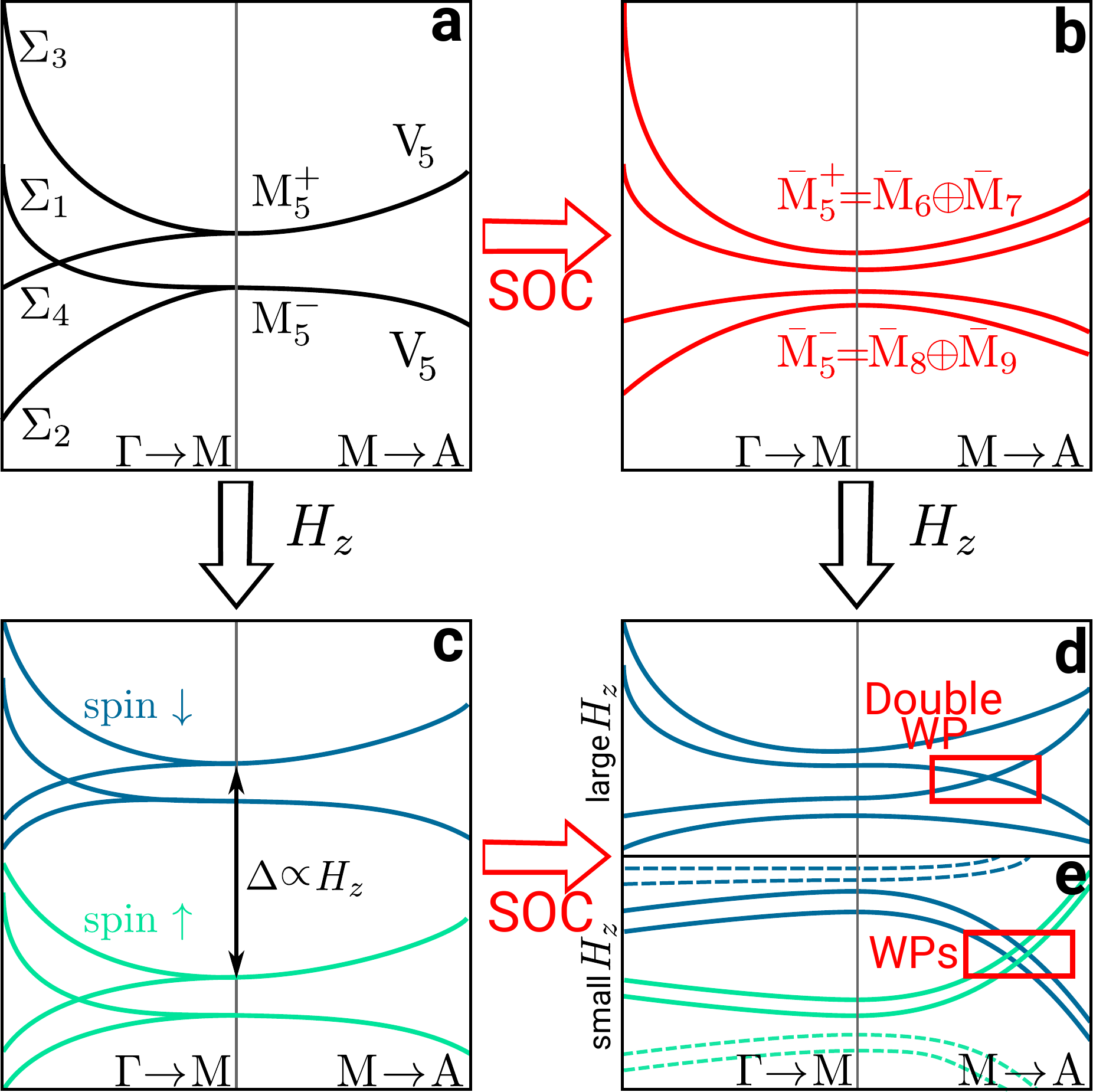}
    \caption{(a) Schematic band structure of CaKFe$_4$As$_4$ along $\Gamma$-$M$ and $M$-$A$. Four bands, transforming under the irreps $\Sigma_{1-4}$, become pairwise degenerate at $M$, where they transform as M$_5^{\pm}$. Along $M$-$A$, they form nodal lines (transforming as V$_5$) which, due to Kramers degeneracy, are fourfold degenerate. (b) In the presence of SOC, the bands transform as the double-valued reducible representations $\bar{\text{M}}_5^\pm$. The two nodal lines are split into four bands that remain Kramers degenerate. (c) In the presence of a magnetic field (but no SOC), the Kramers degeneracy is split into a spin-$\uparrow$ and a spin-$\downarrow$ branch. The Zeeman splitting $\Delta$ is proportional to the field $H_z$. (d)-(e) In the presence of both SOC and a magnetic field, the bands become non-degenerate, and an accidental intersection can occur, resulting in a Weyl point (WP). Its topological charge $\mathcal{C}$ depends on whether the crossing involves bands of the same spin ($\mathcal{C} = \pm 2$, resulting in a double WP, panel (d)) or opposite spins ($\mathcal{C} = \pm 1$, resulting in a single WP, panel (e)).}
    \label{fig:schematic}
\end{figure}

We consider first the case without magnetic field or SOC, where all bands are spin-degenerate. As shown schematically in Fig.~\ref{fig:schematic}(a), which is based on our density functional theory (DFT) calculations, the band structure along the $\Gamma$-$M$ line consists of four bands that transform as the single-valued
representations $\Sigma_{1-4}$ ~\cite{aroyo2011,Supplementary}. Due to the $C_4^z$
symmetry at $M$, these bands become pairwise degenerate at $M$, transforming under the two-dimensional
irreducible representations (irreps) M$_5^\pm$. Microscopically, the energy splitting between the M$_5^\pm$ energy levels is set by the strength of the coupling between the two FeAs layers.
Both the M$_5^+$ and the M$_5^-$ irreps have basis functions with $C_4^z$
eigenvalues $\mp i$, which are related to the orbital moments of the wave-functions. Because the $C_4^z$ axis does not cross the Fe atoms, the orbital moments are not those of the Fe atomic orbitals. Instead, they are determined by the relative phase of the wave-functions on the two different Fe sites on the same plane in the unit cell. 
Continuing along the $M$-$A$ line in Fig.~\ref{fig:schematic}(a), the bands remain pairwise degenerate, transforming as V$_5$. The total degeneracy of each of these bands is four, due to the
Kramers degeneracy enforced by the composition of inversion
and time-reversal symmetries. 

There are two different ways to break this fourfold degeneracy. One is to turn on the SOC, as shown in Fig.~\ref{fig:schematic}(b). In this case, we must consider the symmetries of the double-valued space group, which explicitly accounts for the spin-1/2 of the electrons. Because at the $M$ point the double group has no four-dimensional
representations, the two fourfold degenerate bands split in four
twofold degenerate bands. This remaining twofold degeneracy is a consequence of the fact that the Kramers degeneracy is preserved by SOC. 

The bands at $M$ transform as the
two-dimensional pseudo-real irreps $\bar{\mathrm{M}}_6$, $\bar{\mathrm{M}}_7$, $\bar{\mathrm{M}}_8$, and $\bar{\mathrm{M}}_9$ of the double-group, whose $C_4^z$ eigenvalues are either e$^{\mp i 3\pi/4}$ (for $\bar{\mathrm{M}}_{6,8}$) or e$^{\mp i \pi/4}$ (for $\bar{\mathrm{M}}_{7,9}$)~\cite{Supplementary}. The extra factor of e$^{\mp i \pi/4}$ with respect to the original $C_4^z$ eigenvalues of M$_5^\pm$ originates from the spin angular momentum, such that the eigenvalues in the double-group correspond to the $z$ component of the total angular momentum $(\mathbf{L}+\mathbf{S})$ of the electron.

An alternative way to break the fourfold degeneracy of the $M$-$A$ bands in Fig. \ref{fig:schematic}(a) is to break time-reversal symmetry by applying
an external magnetic field parallel to the $z$-axis,
as indicated in Fig. ~\ref{fig:schematic}(c).
In this case, each band is split into two spin branches, denoted by
spin-$\uparrow$ and spin-$\downarrow$. Their energy splitting $\Delta$
is proportional to the magnetic field $H_z$. The reason why the $M$-$A$ bands retain a twofold degeneracy, even though the Kramers degeneracy is absent, is because of the non-Abelian nature of the little group  on the $M$-$A$ line, which is isomorphic to the magnetic space group $P42'2'$ (\#89.90).

Combining both SOC and $H_z$, all bands become non-degenerate.
Depending on the values of the bilayer splitting between M$_5^\pm$ [Fig. \ref{fig:schematic}(a)], the SOC splitting between $\bar{\mathrm{M}}_{6,7}$ [Fig. \ref{fig:schematic}(b)], and the Zeeman splitting $\Delta$ between the spin-$\uparrow$ and spin-$\downarrow$ branches [Fig. \ref{fig:schematic}(c)], 
crossings of these non-degenerate bands may occur along the $M$-$A$ line, resulting in the Weyl points shown in Figs. \ref{fig:schematic}(d)-(e). Of course, among these three parameters, the only one that can be efficiently tuned experimentally is $\Delta$. Although group theory cannot predict the existence and positions of these crossings, it can be employed to identify two different scenarios. When $\Delta$ is large compared to the SOC and bilayer splittings, crossings between states with the \emph{same spin orientation} and opposite orbital angular momentum can occur, as shown in Fig. \ref{fig:schematic}(d).
Because, as explained above, states of opposite orbital angular momentum have $C_4^z$ eigenvalues $\pm i$, their ratio is equal to $-1$, resulting in double-Weyl points with topological charge $\mathcal{C} = \pm 2$~\cite{fang2012multi}. On the other hand, when $\Delta$ is small compared to the SOC and bilayer splittings, the crossing can involve states with \emph{opposite spin states} but the same orbital angular momentum, as illustrated in Fig. \ref{fig:schematic}(e). In this case, the ratio between the $C^z_4$ eigenvalues of the crossing bands is equal to $\pm i$, yielding a single-Weyl point with $\mathcal{C} = \pm 1$. In either scenario, note that the Weyl points are anchored in the $C^z_4$ symmetry of the crystal.


To verify whether Weyl points emerge in CaKFe$_4$As$_4$, in Fig.~\ref{fig:cell_and_bands}(b) we present relativistic DFT calculations of the electronic band structure obtained with the Full Potential Local Orbital (FPLO) code \cite{koepernik1999full} within the generalized gradient approximation (GGA) \cite{perdew92}.
To model the effects of an external magnetic field, we assume a ferromagnetic configuration for
the Fe atoms, converging the structure self-consistently
with a fixed total magnetic moment. 
We initiate the calculations with magnetic moments pointing along the [001]
direction, 
resulting in an out-of-plane
magnetic moment of $\mu_{\mathrm{Fe}} = 0.63\mu_B$ on the Fe atoms and $\mu_{\mathrm{As}} =
-0.13\mu_B$ on the As atoms. 
The upper inset
of Fig.~\ref{fig:cell_and_bands}(b) shows a crossing between two $d_{x^2-y^2}$ bands in the spin-$\downarrow$ branch. We thus identify a pair of Weyl points located at $\mathbf{k}_0^\pm = (1/2, 1/2, \pm0.0178)2\pi/a$ and 0.27eV below the Fermi
level.
On the other hand, the spin-$\uparrow$ branch, which
is pushed down in energy by the Zeeman splitting, does not exhibit any band crossing. 
We note that due to the distinct locations of the spin-$\uparrow$ and -$\downarrow$ branches in the energy spectrum, the bands hybridize differently with surrounding As $4p$-bands.

\begin{figure}
    \centering
    \includegraphics[width=\linewidth]{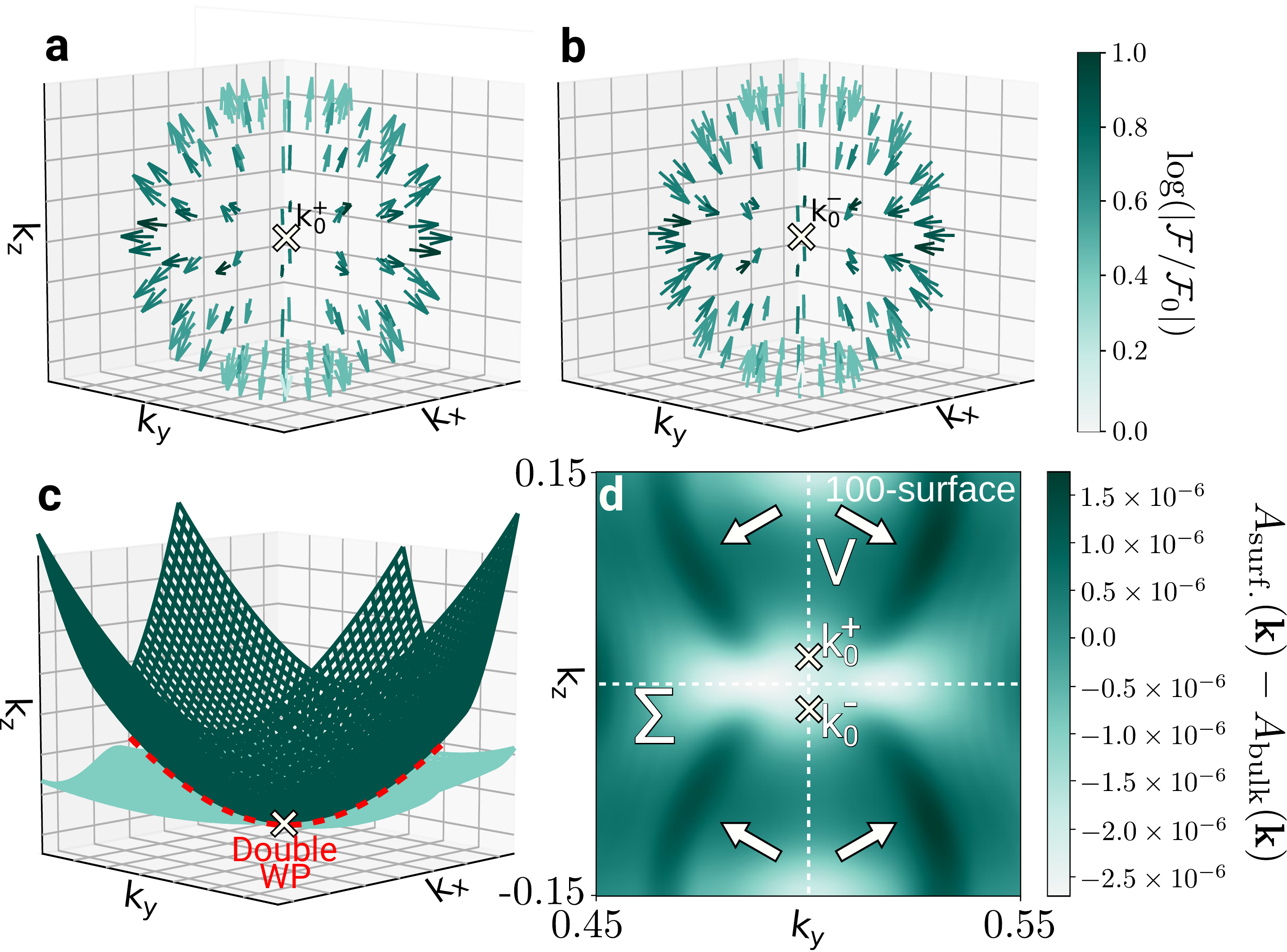}
    \caption{Berry curvature flux $\mathcal{F}$ on momentum-space spheres of radius
	$0.002 \pi/a$ centered at (a) $\mathbf{k}_0^+$ and (b)
	$\mathbf{k}_0^-$. The vectors have been normalized and their magnitudes
	are indicated by their color. Upon integration of $\mathcal{F}$, we obtain
	the topological charges $\mathcal{C}=\pm2$ characteristic of double-Weyl points. (c) Energy dispersion of the two bands forming the Weyl point near $\mathbf{k}_0^+$. Although they intersect
	linearly along $k_z$, a quadratic touching is realized in the
	$(k_x, k_y)$-plane. (d) Spectral density $A(\mathbf{k})$ calculated from DFT across the
	(100) surface. The Fermi arcs are indicated
	by white arrows. In order to enhance the contrast, we performed 
	semi-slab and bulk calculations and subtracted the latter
	from the former.}
    \label{fig:topology}
\end{figure}

As expected from the above symmetry analysis, by expanding the band dispersion in the vicinity of
the two Weyl points at $\mathbf{k}_0^\pm$, we find that they disperse linearly along the $k_z$ direction but quadratically in the $(k_x,k_y)$-plane (see Fig.~\ref{fig:topology}(c)). Thus, the corresponding $\mathbf{k}\cdot\mathbf{p}$ Hamiltonian~\cite{dantas2020non} is:
\begin{equation}
   \mathcal{H} =  m\sigma_z + (ak_+^2 + bk_-^2)\sigma_+ + h.c.,
\end{equation}
with $k_\pm = k_x \pm i k_y$ and Pauli matrices $\sigma_i$ defined in the subspace of the two bands. 
As a result, a double
Weyl point with topological charge $\mathcal{C} = \pm 2$ is realized. 
This is verified by directly computing the Berry curvature flux across two small spheres enclosing the Weyl points from our DFT calculations. As shown in Figs.~\ref{fig:topology}(a)-(b), the two nodes act as a source and a sink of Berry flux $\mathcal{F}$, represented by the vector field. Integrating the Berry flux numerically, 
we indeed obtain topological charges $\mathcal{C} = \pm2$. This non-trivial topological charge is expected to manifest itself experimentally as an anomalous Hall effect (AHE)~\cite{Nagaosa06,Nagaosa10}.

Another typical manifestation of Weyl fermions is the emergence of Fermi arcs on the sample's surfaces. In Fig.~\ref{fig:topology}(d), we show the $k$-resolved spectral density on the
(100) surface at the energy where the Weyl points are located. Two pronounced
Fermi arcs are identified, as indicated by the white arrows, terminating at the two Weyl points, denoted by the crosses. 

To experimentally observe these manifestations of the Weyl fermions -- AHE and Fermi arcs -- it is desirable that the Weyl points are close to the Fermi level $E_F$. In our DFT calculations, they are located $0.27$ eV below $E_F$. However, it is well-established that DFT overestimates the energies of the bottom/top of the bands, which are reduced by correlations \cite{Ortenzi2009,yin2011kinetic,ferber2012,Chubukov2016,zantout2019,Bhattacharyya2020}. Indeed, ARPES experiments in CaKFe$_4$As$_4$ found that the energies of the bottom of the electron pockets were at least five times smaller than those predicted by DFT \cite{Kaminski2016}. This suggests that the Weyl points in the actual compound are likely much closer to $E_F$. Moreover, hole doping could be used to further tune the Fermi energy to the desired position.

In summary, we demonstrated that the breaking of the Fe-plane inversion symmetry in iron-based superconductors, combined with an external magnetic field, provide a promising route to realize Weyl points in these materials. While our focus here was on CaKFe$_4$As$_4$, the general symmetry arguments are expected to apply to any other iron-based material whose Fe layers lack inversion symmetry. The latter is expected to happen, for instance, in thin films of single-layer compounds~\cite{Hao2014Topological}, since the substrate explicitly breaks the inversion symmetry. Alternatively, it can take place as a spontaneous symmetry-breaking driven by interactions that favor the spin-vortex crystal magnetic ground state, which was recently observed in the phase diagram Ba$_{1-x}$Na$_x$Fe$_2$As$_2$~\cite{Sheveleva2020muon}.

More broadly, the realization of Weyl points in iron-based materials would open the door to investigate the interplay between Weyl fermions and other types of electronic orders not usually present in the currently studied Weyl semimetals. For instance, spontaneous nematic order or applied uniaxial strain would break the $C_4^z$ symmetry and  split the double-Weyl points into two single-Weyl points (see \cite{Supplementary} for more details). Another interesting route would be to probe the impact of unconventional superconductivity, which onsets below $30$K in CaKFe$_4$As$_4$, on the Weyl fermions and the associated Fermi arcs.

\begin{acknowledgments}

The authors acknowledge fruitful discussions with 
S. M. Winter, K. Kopernik and J. L. Mañes. NH and RV were financially supported by the Stiftung Polytechnische Gesellschaft Frankfurt and by the Deutsche Forschungsgemeinschaft
(DFG, German Research Foundation) through TRR 288
- 422213477 (project A05, B05). MHC and RMF were supported by the the U. S. Department of Energy, Office of Science, Basic Energy Sciences, Materials Sciences and Engineering Division, under Award No. DE-SC0020045. MHC acknowledges support by the Villum foundation during the writing of this manuscript. S-S.Z. and C.D.B. are supported by funding from the Lincoln Chair of Excellence in Physics.
TB and FY were supported by the Office of Naval Research Grant N00014-20-1-2361.
MI acknowledges support from the  Spanish  Ministerio  de  Ciencia  e  Innovacion (grants number PID2019-109905GB-C21 and PGC2018-094626-B-C21) and Basque Government (grant IT979-16). CB, RV and RF thank the Kavli Institute of Theoretical Physics (KITP), which is
supported by  the National Science Foundation under Grant No.NSF PHY-1748958.

\end{acknowledgments}

\bibliography{refs}

\cleardoublepage


\section*{Supplementary material}

# Prediction of Double-Weyl Points in the Iron-Based Superconductor CaKFe$_4$As$_4$: Supplementary Material

Niclas Heinsdorf,[1] Morten H. Christensen,[2,][*] Mikel Iraola,[1,3,4] Shang-Shun Zhang,[5] Fan Yang,[6] Turan Birol,[6] Cristian D. Batista,[5,7] Roser Valentí,[1] and Rafael M. Fernandes[2]

[1]Institut für Theoretische Physik, Goethe-Universität Frankfurt, Max-von-Laue-Strasse 1, 60438 Frankfurt am Main, Germany
[2]School of Physics and Astronomy, University of Minnesota, Minneapolis, 55455 MN
[3]Department of Condensed Matter Physics, University of the Basque Country UPV/EHU, Apartado 644, 48080 Bilbao, Spain
[4]Donostia International Physics Center, 20018 Donostia-San Sebastian, Spain
[5]Department of Physics and Astronomy, The University of Tennessee, Knoxville, Tennessee 37996, USA
[6]Department of Chemical Engineering and Materials Science, University of Minnesota, Minneapolis, 55455 MN
[7]Neutron Scattering Division and Shull-Wollan Center, Oak Ridge National Laboratory, Oak Ridge, Tennessee 37831, USA

| $\Sigma$ | | $V$ | |
|---|---|---|---|
| $M_1^-(1) \to \Sigma_4(1)$ | | $M_1^-(1) \to V_4(1)$ | $A_1^-(1) \to V_4(1)$ |
| $M_3^+(1) \to \Sigma_2(1)$ | | $M_3^+(1) \to V_4(1)$ | $A_3^+(1) \to V_4(1)$ |
| $M_4^-(1) \to \Sigma_4(1)$ | | $M_4^-(1) \to V_2(1)$ | $A_4^-(1) \to V_2(1)$ |
| $M_2^+(1) \to \Sigma_2(1)$ | | $M_2^+(1) \to V_2(1)$ | $A_2^+(1) \to V_2(1)$ |
| $M_5^+(2) \to \Sigma_3(1) \oplus \Sigma_4(1)$ | | $M_5^+(2) \to V_5(2)$ | $A_5^+(2) \to V_5(2)$ |
| $M_5^-(2) \to \Sigma_1(1) \oplus \Sigma_2(1)$ | | $M_5^-(2) \to V_5(2)$ | $A_5^-(2) \to V_5(2)$ |

TABLE I. Compatibility relations along $M$-$\Sigma$ and $M$-$V$ for the single-valued space group $P4/mmm$ [1].

| $M$ | $A$ |
|---|---|
| $\bar{M}_6 \to \bar{V}_6$ | $\bar{A}_6 \to \bar{V}_6$ |
| $\bar{M}_7 \to \bar{V}_7$ | $\bar{A}_7 \to \bar{V}_7$ |
| $\bar{M}_8 \to \bar{V}_6$ | $\bar{A}_8 \to \bar{V}_6$ |
| $\bar{M}_9 \to \bar{V}_7$ | $\bar{A}_9 \to \bar{V}_7$ |

TABLE II. Compatibility relations of the double-valued space group $P4/mmm$.

## COMPATIBILITY RELATIONS

In Tables I and II we show the compatibility relations along the $M$-$\Sigma$ and $M$-$V$ lines for the single-valued and double-valued space group $P4/mmm$ of CaKFe$_4$As$_4$.

## METHODS

We computed the electronic structures using density functional theory (DFT) within the Perdew-Burke-Ernzerhof parameterization of the generalized-gradient approximation for the exchange-correlation functional[2, 3] as available in the Full Potential Local Orbital (FPLO) code [4–9] and the Vienna-ab-Initio package [10–12]. The representations under which the Bloch states transform have been identified using the 'irrep' code provided with data from the Bilbao Crystallographic Server (BCS) [1]. Experimental lattice constants were employed [13] and the electron density has been converged on a $17 \times 17 \times 17$ k-point mesh. Spin-orbit effects were included self-consistently. To calculate the symmetry eigenvalues and the topological invariants, a tight-binding model containing the Fe 3$d$- and the As 4$p$-orbitals was constructed. The hopping amplitudes were obtained by projection on the corresponding Wannier functions [14], using the FPLO WF-module (within an energy window of -6 to 2 eV and Gaussian tails of 0.5 standard deviation).

## CRYSTAL STRUCTURE AND BAND STRUCTURE CALCULATIONS

CaKFe$_4$As$_4$ consists of layers of FeAs$_4$ tetrahedra arranged in a square-lattice fashion separated by alternate layers of Ca and K (see Fig. S1(a)). The material crystallizes in the symmorphic tetragonal space group $P4/mmm$ (#123) with lattice constants $a = b = 3.8659\text{Å}$ and $c = 12.884\text{Å}$. It is inversion symmetric and features a set of mirror planes as well as fourfold rotations about the $z$-axis $C_4^z$ centered at either the middle of the conventional unit cell or at one of its four corners. Contrary to the parent $A$Fe$_2$As$_2$ systems with $A$ alkali or alkaline-earth cations, the glide plane symmetry is broken due to the inequivalence of Ca and K (see Fig. S1(a)). The electronic structure of CaKFe$_4$As$_4$ around the Fermi energy, as calculated from density functional theory (DFT) within the generalized gradient approximation (GGA), consists of iron 3$d$-bands hybridizing with arsenic 4$p$-bands (see Fig. S1(b)). The strong interlayer As 4$p_z$ coupling results in large-bandwidth $p_z$-bands that intersect with the iron 3$d$ bands. These crossings are protected by inversion symmetry and experience (with inclusion of spin-orbit coupling) band inversions as well as topological Dirac semimetal crossings, as has been recently investigated elsewhere [15]. In the manuscript, we concentrate on crossings of non-degenerate iron 3$d$-bands instead, and on the emergence of possible time-reversal symmetry-broken Weyl points.

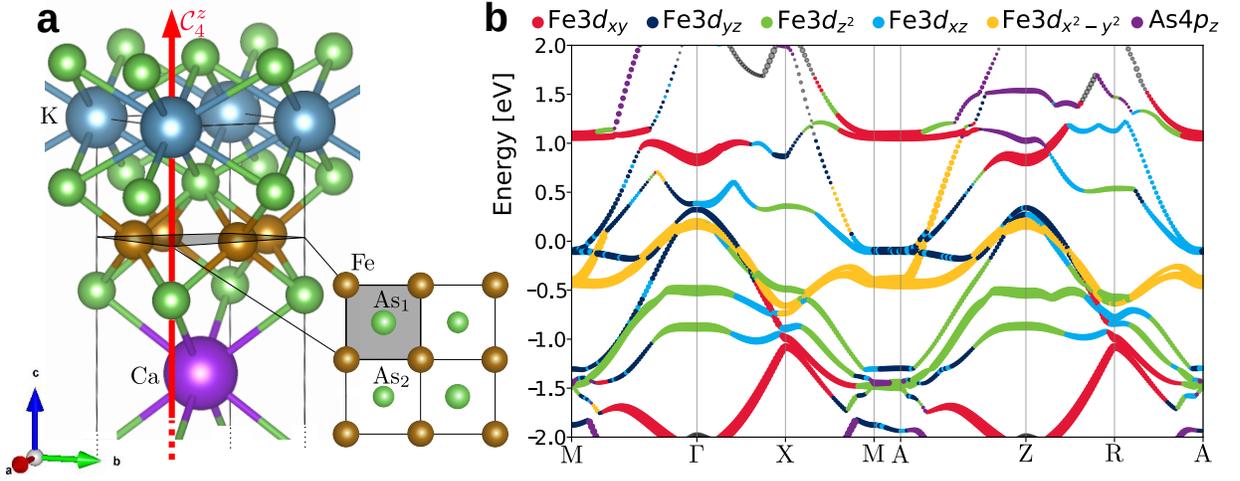

FIG. S1. (a) Crystal structure of CaKFe$_4$As$_4$ with a top view on the upper FeAs layer (gray). The system is invariant under fourfold rotations with respect to an axis centered at the edges of the tetragonal unit cell and parallel to the $z$-axis, as indicated by the red arrow. Additionally, the system remains invariant under reflections with respect to any of the unit cell's boundaries, or to a mirror placed at the center of the unit cell, intersecting the Ca atom, with a normal parallel to the $z$-axis. The system is inversion symmetric, but in contrast to many other iron-based pnictides, there is no glide-plane symmetry, which is broken by the inequivalence of the interlayer atoms, namely K and Ca. (b) DFT band structure of paramagnetic CaKFe$_4$As$_4$. For each data point the dominating orbital character is given by its color, with gray corresponding to an orbital not listed in the legend above. The thickness of the bands displays the magnitude of the respective orbital's projection onto the Kohn-Sham state.

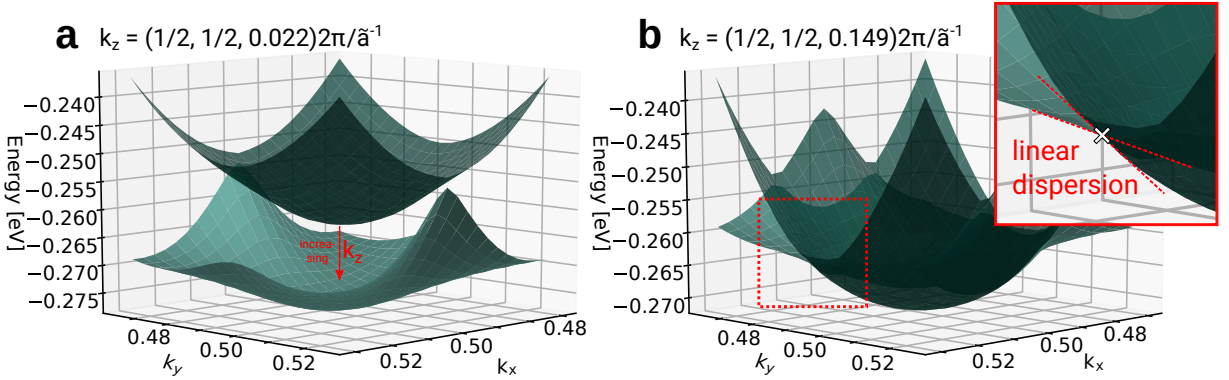

FIG. S2. Energy dispersion in the $(k_x, k_y)$-plane at (a) $k_z = (1/2, 0.494, 0.022)2\pi/\tilde{a}^{-1}$ and (b) $k_z = (1/2, 0.494, 0.149)2\pi/\tilde{a}^{-1}$ for CaKFe$_4$As$_4$ with an $a$-axis compressed by 1%. The fourfold rotational symmetry $C_4^z$ along V that protects the higher order chirality of the Weyl points on this line is broken by the compression. As a consequence, the double Weyl point is split, and the resulting single Weyl points are shifted away from $M$ into the $(k_x, k_y)$-plane. In their immediate vicinity the dispersion is linear, in contrast to the quadratic dispersion of the double Weyl points shown in the main text.

## $C_4^z$ BREAKING

In the main text we argued that the double Weyl point is stabilized by the fourfold rotational symmetry $C_4^z$ along the $V$-line in the first Brillouin zone. This symmetry can be broken by compression along the $a$-direction, which squeezes the unit cell vector by 1% to $\tilde{a} = 3.82724 \text{Å}$. In Fig. S2 we show the dispersion around the $M$ point in the $(k_x, k_y)$-plane for different, fixed values of $k_z$ in the case where the $C_4^z$ symmetry is broken. Just as for the uncompressed material, the distance between the two bands becomes smaller for larger crystal momentum $k_z$, until the two bands intersect. As a result of the broken point group symmetry $C_4^z$, the new Weyl points are shifted away from $M$ into the $(k_x, k_y)$-plane. Without $C_4^z$ there is no longer a constraint on the order of the dispersion in the proximity of the crossing. Zooming in on the crossing reveals the linear dispersion relation of a single-Weyl point.

---

* Present Address: Center for Quantum Devices, Niels Bohr Institute, University of Copenhagen, Copenhagen

2100, Denmark



\end{document}